# Phase diagrams of polymer-containing liquid mixtures with a theory-embedded neural network

Issei Nakamura[1]

[1] Department of Physics, Michigan Technological University, Houghton, Michigan 49931, United States

E-mail: inakamur@mtu.edu

**Abstract**

We develop a deep neural network (DNN) that accounts for the phase behaviors of polymer-containing liquid mixtures. The key component in the DNN consists of a theory-embedded layer that captures the characteristic features of the phase behavior via coarse-grained mean-field theory and scaling laws and substantially enhances the accuracy of the DNN. Moreover, this layer enables us to reduce the size of the DNN for the phase diagrams of the mixtures. This study also presents the predictive power of the DNN for the phase behaviors of polymer solutions and salt-free and salt-doped diblock copolymer melts.

## 1. Introduction

   The phase separation of polymer-containing liquid mixtures is ubiquitous over a broad spectrum of science and engineering. It is necessary to analyze the abundant and diverse experimental data obtained concurrently; however, this process may be a daunting task when experimental settings and material properties are altered, and simulations may not assess the relevant time scale of phase separation. Furthermore, accurate correspondence among an experiment, theory, and simulation often becomes unclear.
   Among others, as in many other problems in physics, the study of phase separation and drawing phase diagrams often faces the following challenges: (1) (Statistical) thermodynamic theories may be constructed to account for the transition between two phases observed in experiments. However, solving equations derived from the theories often becomes a computationally intractable or daunting task. For example, molecular theories in statistical thermodynamics, such as self-consistent mean-field theory, integral-equation theory, and density-functional theory, are often invoked to consider liquid-liquid phase separation [1,2]. However, various situations can easily exponentially increase the complexity of the equations in the theories. Examples of the relevant systems include highly branched polymers and semi-flexible polymers, to name a few. Thus, computationally tractable models that encompass or approximate the theories are greatly needed. (2) It is often challenging to construct a molecular theory that accounts for experimental data because the complete mechanism of the phase separation is unclear. For example, block copolymer electrolytes [3,4] and nanoparticle-containing polymers [5-7] often fall in this class of system. In this case, we may construct phenomenological models, which are expected to maximally involve underlying thermodynamic features in experimental data. (3) Theories can be available, but experimental data for the phase diagram of the phase separation are limited. To examine the theories, we must extrapolate and/or interpolate the experimental data. Indeed, small-angle X-ray scatterings and small-angle neutron scatterings are typically performed to analyze the microphase separation of salt-doped block copolymers, but the overall phase diagrams are deduced from the linear interpolation of the limited number of observed data points [8-10]. However, it is often impractical to interpolate the phase boundaries according to underlying thermodynamic features in the targeted systems. Typically, molecular interactions are significantly correlated through the chain connectivity of polymers, thus substantially increasing difficulty with coping with the three situations. This fact appears to suggest that a more practical approach to addressing these situations regarding the phase behavior may be constructed by weaving model-independent methods, but the efficacy and feasibility of the approach should also be judged by the computational effort required in modeling and analysis.

In this article, we present a relatively model-independent approach to addressing the three issues by developing a deep neural network (DNN) for the phase separation of both salt-free and salt-doped polymer-containing liquids. In the current study, the DNN does not require many neurons and hidden layers and is thus computationally feasible on common workstations. Note that convolutional neural networks consist of convolutional and pooling layers for extracting the characteristic features of visual imagery and reducing the dimensions of the input data [11]. Inspired by this perspective, we construct the first layer via coarse-grained mean-field theory and scaling laws. This layer captures the representative features of target systems, such as the entropy and enthalpy of polymer mixtures and the dependence of the domain spacing of block copolymer melts on salt concentrations. This layer also allows us to ease local minimum problems in loss functions that evaluate the accuracy of NNs [11]. The output of our DNNs is given by a Gaussian function whose values are assigned to phase behavior types, such as the macroscopic phase separation of polymer solutions and the ordered structures of block copolymer melts. In this sense, phase diagrams per se are empirically embedded in the DNNs. Thus, we must know what type of phase separation may occur prior to the calculation of the DNNs, but the DNNs enable easier phase identification from molecular properties.

NNs have drawn considerable attention in solving real-world problems over a broad spectrum of research areas and have already shown their remarkable efficacy in solving practical scientific problems [12-22]. Early examples for molecular separation include the study by Dai, Sumpter and Noid, who developed an early type of NN that accounted for the phase boundary of the macroscopic phase separation of binary molten-salt mixtures [23]. Later, NNs were used to consider the phase diagrams for microemulsion-based drug delivery systems involving oil, water and surfactants [24]. The architecture of the NN was relatively simpler than those of the recent DNNs, yet the NN, tuned with only 171 training data points, became remarkably consistent with the experimental pseudo-ternary phase diagrams. Several studies related to microemulsions for drug delivery systems further provided the proof-of-concept that NNs may have excellent predictive power for the phase behavior of liquid mixtures [25-30]. Nevertheless, the development of NNs for the phase behavior of polymer-containing liquid mixtures remains significantly limited. Part of the reason appears to arise from some technical requirements that appear to substantially restrict the use of NNs by non-experts in machine learning. For example, NNs are typically trained by a gradient-descent optimization algorithm such as a backpropagation algorithm using package code or computer software. However, loss functions may involve many local minima, which substantially declines the efficiency of the backpropagation method because the gradient descent gets trapped in local minima. Moreover, the landscape of the loss function often has many saddle points, at which the gradient becomes zero. This feature also significantly slows down the update process of model parameters (weights) when the backpropagation algorithm is invoked. Additionally, the initial guess of the model parameters is also often critical in training NNs. In this study, to circumvent or minimize these technical concerns and implement relatively effortless and extensible programming, we perform an alternative approach in which we randomly search the weights without backpropagation algorithms. We show that the first layer, in which thermodynamic theory is embedded, substantially enhances the efficiency of the random searches and increases the accuracy of the DNNs.

**2. Neural network for polymer solutions**

Our DNN for both polymer solutions and diblock copolymer melts consists of three layers (Figure 1). In this study, we use three, ten, and two neurons in hidden layers 1, 2, and 3, respectively. For the input $x$ in hidden layers 2 and 3, we use the rectified linear unit (ReLU) $\text{ReLU}(\sum w_{ij} x_i + b_j)$ for activation functions with weight $w_{ij}$ and bias $b_j$. The key component in the DNN is the theory-embedded layer (hidden layer 1), which captures the significance of physical properties and thus "speculates" the characteristic features of the phase behaviors. Here, the Flory-Huggins theory for polymer solutions suggests that (a) the overall trend of spinodal curves for phase instability is substantially affected by the difference in the translational entropy between polymers and solvents, (b) the chain length of polymers (or the degree of polymerization) plays a crucial role in producing asymmetry in the shapes of spinodal curves and thus in determining the location of the critical point, and (c) the phase boundary is determined by a delicate balance between the entropy and enthalpy of polymers and solvents. Accordingly, we cast these three pieces of information into nonlinear functions:

$$f_1 = \phi \ln(\phi) / [(1 - \phi) \ln(1 - \phi)]$$

$$f_2 = 1/N$$

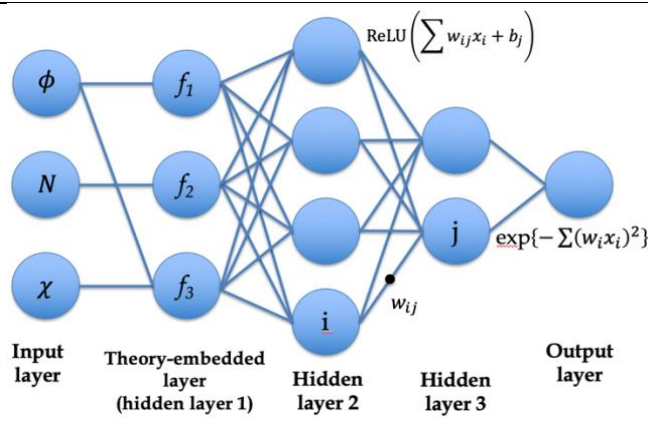

**Figure 1.** Deep neural network for the phase behavior of polymer-containing liquid mixtures. The theory-embedded layer (hidden layer 1) captures the significance of physical properties. This layer may also be fully connected to the input layer. Hidden layers 2 and 3 serve as standard layers.

$$f_3 = \chi\phi(1-\phi) \tag{1}$$

where $\phi$, $N$, and $\chi$ designate the volume fraction of polymers, the chain length of polymers, and the Flory parameter for measuring immiscibility between species, respectively. $f_1$ and $f_3$ calculate the ratio of the translational entropies of incompressible polymers and solvents and the enthalpic contribution to the free energy, respectively. Given that $M$ is the number of training data points, a loss function is defined as the relative error $\sum \sigma_i/M$. Here, $\sigma_i = 1$ if the estimated and true values are different; otherwise, $\sigma_i = 0$. We minimize the loss function by randomly searching the weights [31]. To examine the efficacy of the theory-embedded layer, we consider the following patterns: $f_1$ is replaced by $\phi \ln(\phi)$ (pattern 1) and $f_1 = \phi$, $f_2 = N$ and $f_3 = \chi$ (pattern 2). We terminate a random search if the relative error is larger than 8% when $2 \times 10^6$ trials are performed. The success rate of the intact form is estimated at 8.7% from more than 20000 independent simulation runs for the same training datasets, but that of pattern 1 and pattern 2 is 0%. Thus, pattern 1 and pattern 2 are unsuccessful because they do not account for information about the balance between the entropies of polymers and solvents.

Note that standard neurons with activation functions transmit a signal to other neurons when the strength of the signal exceeds a certain threshold. Thus, the mean-field theory incorporated in the theory-embedded layer evaluates the qualities of the physical properties and transforms them into signals for hidden layer 2. If the signals are significant enough to exceed the threshold in the activation functions, then they will be further transmitted to other neurons.

The output layer determines the phase behavior of polymer solutions by calculating a Gaussian function,

$$y = \exp\left\{-\sum_i (w_i x_i)^2\right\} \tag{2}$$

Here, a disordered (DIS) phase is assigned when the output $y < 0.5$; otherwise, macroscopic phase separation is assigned. Note that we obtain macroscopic phase separation when all weights $w_i$ are zero. Thus, gradually increasing the weights $w_i$ from zero indicates that the DNN initially fits the training data points for macroscopic phase separation and then gradually tunes the weights for DIS. Indeed, this fact provides a computational advantage in performing a random search for the weights $w_i$. Thus, we generate the initial values of $w_i$ from -1 to 1 and then perform random searches by calculating $w_i \rightarrow w_i \pm \Delta$, where $\Delta$ is a uniform random number between -1 and 1. If the relative errors are large after a certain number of trials, then we discard the trial process and generate another initial value. This approach is used because training datasets often involve the deep local minima of the loss function. However, if the initial values are appropriately set, then a random search can be finished quickly; for example, in certain cases, the search can be finished even within a few seconds. Here, the current study employed standard CPU cores (Intel Sandy Bridge E5-2670 2.60 GHz or equivalent). The theory-embedded layer allows us to promptly reach relative errors of less than 5%. The computational time for a single run typically ranges from a few seconds to no more than a few hours throughout this article. Thus, our random search with the Gaussian function is computationally feasible on standard workstations.

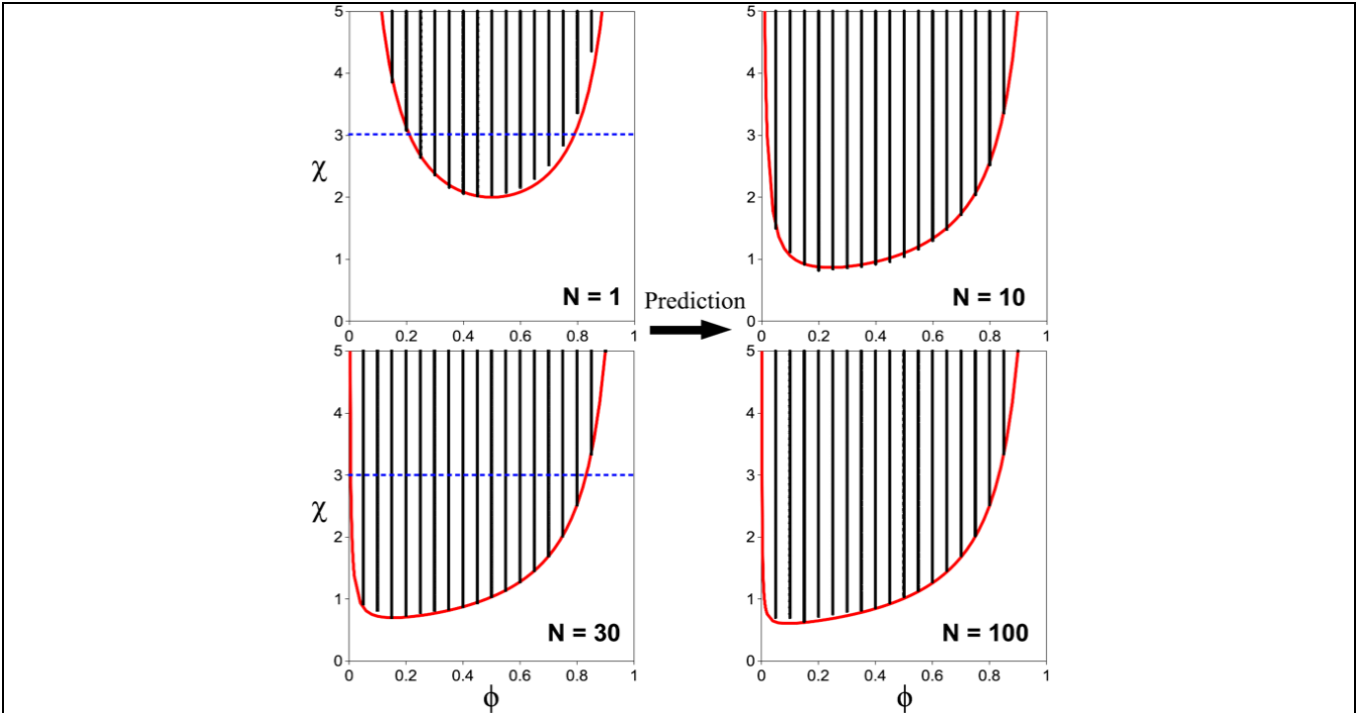

**Figure 2.** Phase diagrams of polymer solutions. The training data points exist below $\chi = 3$ (blue dashed lines in the left two figures). The DNN generates the black regions for macroscopic phase separation, which consist of dense data points that form "strip" structures in the y-direction. The Flory-Huggins theory provides the spinodal curves (red solid lines), above which macroscopic phase separation occurs. The DNN predicts the right two figures.

We provide the training and test datasets for the phase behavior using the Flory-Huggins theory as follows: We calculate the free energy $F = \frac{\phi}{N}\ln(\phi) + (1-\phi)\ln(1-\phi) + \chi\phi(1-\phi)$ and collect the information of whether polymer solutions phase separate or not (i.e., a "yes/no" data type) by calculating spinodal curves using $F$ [32,33]. The first, second, and third terms account for the translational entropies of polymers and solvents and the intermolecular interaction, respectively. The intervals of $\phi$ and $\chi$ are 0.05, and $\phi$ and $\chi$ range from 0 to 1 and 0 to 3, respectively. Our rationale for examining the DNNs with the Flory-Huggins theory is that the theory can capture the shift in the critical point in the phase diagrams as the chain length $N$ of polymers is increased. Accordingly, the spinodal curves become more asymmetric as $N$ is increased. This characteristic feature is indeed observed in experiments for polymer solutions, and we suggest that whether the DNNs can capture this characteristic feature should provide a benchmark for the efficacy of the DNNs. The chain lengths are $N = 1$ for nonpolymeric liquids and 30 for polymer solutions. Figure 2 shows the phase diagrams produced by the DNN. The relative error is 0.094%. The agreement between the predictions of the DNN and the test datasets is excellent within $\chi \leq 3$. The DNN also predicts the phase behaviors for $N = 1$ and 30 in the range of $3 < \chi \leq 5$, and the results are in excellent agreement with the test datasets. Moreover, the DNN predicts the phase behaviors for $N = 10$ and 100 remarkably accurately.

## 3. Neural network for salt-free block copolymer melts

We can also directly apply our DNN shown in Figure 1 to the phase behavior of incompressible, symmetric diblock copolymer melts. The inputs are the volume fraction of block A $\phi$, the chain lengths of blocks A and B $N$, and the Flory parameter $\chi$. Thus, no essential changes in the DNN for polymer solutions are necessary. The value $y$ of the Gaussian function in the output layer is defined as follows: $0 < y \leq 0.2$ for DIS, $0.2 < y \leq 0.4$ for body-centered-cubic (BCC) phases, $0.4 < y \leq 0.6$ for hexagonally packed cylinder (HEX) phases, $0.6 < y \leq 0.8$ for gyroid (GYR) phases, and $0.8 < y \leq 1$ for lamellar (LAM) phases. We have assigned these ranges heuristically, but we have not encountered any inconvenience in this study. We consider the following characteristic features of the phase behavior of diblock copolymer melts: (a) the translational entropy of each block, (b) the enthalpic interaction between the two blocks, and (c) the interfacial width between the two blocks in the ordered structures. A scaling law suggests that the interfacial width in the ordered structures scales with $\approx (N\chi)^{-0.5}$ for weakly segregated diblock copolymers [34]. Accordingly, we set

$$f_1 = \phi \ln(\phi) + (1-\phi)\ln(1-\phi)$$

$$f_2 = \chi\phi(1-\phi)$$

$$f_3 = (N\chi)^{-0.5} \tag{3}$$

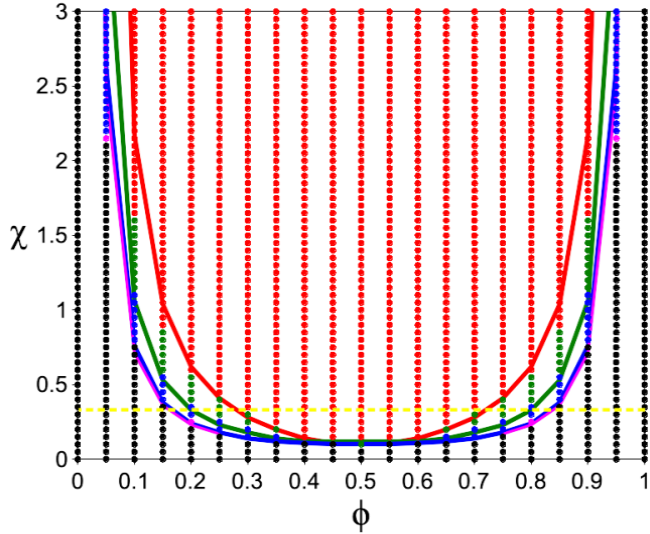

**Figure 3.** Phase diagram of diblock copolymer melts. The training dataset exists below $\chi = 0.3$ (yellow dashed line). The DNN predicts the phase behavior above $\chi = 0.3$. The Landau theory provides the phase boundaries between GYR and LAM (red), HEX and GYR (green), BCC and HEX (blue), and DIS and BCC (purple). Data points: LAM (red), GYR (green), HEX (blue), BCC (purple), and DIS (black).

The optimization performance was not sensitive to the exponent in $f_3$. This study also examined several other functional forms for $f_i$, which did not adequately capture the qualitative features of block copolymer melts, but those functional forms failed to tune the DNN. However, once the $f_i$ suggested above was used, the performance drastically increased, and tuning the weights with high accuracy became achievable. To evaluate the efficacy of $f_i$, a random search was terminated after $5\times10^7$ trials if the relative error was larger than 5%. The success rate was 0.23% out of 3047 trials, whereas that of the other patterns was 0%.

The training and test datasets for the phase behavior were produced using the Landau theory of microphase separation developed by Leibler [35] and Hamley and Podneks [36]. In these theories, we can calculate the free energies for the ordered structures of block copolymer melts using mean-field approximations and draw the phase boundaries of the ordered structures. The theories are well known to qualitatively capture the experimental phase diagrams of diblock copolymer melts. Thus, we suggest that the datasets derived from these theories serve as a good model system to examine our DNNs. The training dataset consists of 365 data points equally spaced in $0 \leq \phi \leq 1$ and $0 \leq \chi \leq 0.3$. The chain length of a block copolymer is $N = 100$. Figure 3 shows the phase diagram drawn by the DNN. The relative error is 3.84%. The predictions of the DNN in $0.3 < \chi \leq 3$ are remarkably consistent with the test dataset. The test dataset shows that BCC exists between the purple and blue lines, but this area is quite narrow. Nevertheless, the DNN also adequately captures this feature and compares favorably with the test dataset, providing purple dots for BCC at, for example, $\phi = 0.05$ and $0.95$. This study examined several other sets of weights with relative errors less than 5%, and the predictive power was equivalently good when $\chi \leq 2$. Thus, the predictions of the DNN are relatively robust in considering the phase behavior of block copolymer melts.

**4. Neural network for salt-doped block copolymer melts**

TABLE I. *Unsuccessful* trial patterns of $f_i$ vs. the success rate. A random search was terminated after $5\times10^7$ trials if the relative error was larger than 10%.

|  | $f_1$ | $f_2$ | $f_3$ | Success rate % |
|---|---|---|---|---|
| Pattern 1 | $r$ | $N$ | $\chi$ | 0 |
| Pattern 2 | $r/N$ | $1/N$ | $r\chi$ | 0 |
| Pattern 3 | $r/N$ | 0 | $r\chi$ | 0 |
| Pattern 4 | $r$ | $1/N$ | $\chi/r$ | 0 |
| Pattern 5 | $r$ | $1/N$ | $r+\chi$ | 0 |
| Pattern 6 | $\ln r$ | $15r$ | $1/\sqrt{N(r+\chi)}$ | 0.03 |

We also examine the efficacy and predictive power of our DNN for experiments, taking the phase behavior of lithium salt-doped PEO-b-PS diblock copolymer melts as an example [9]. This system exhibits disorder-order and various order-order phase transitions. However, reproducing the experimental observations and understanding the true nature of the phase behaviors by existing theories and molecular simulations remain significantly limited, primarily because both computational and theoretical modeling and calculations are challenging. Note that the microphase separation of this system can be characterized by (a) the translational entropy of the salt ions $r\ln r$, (b) the enthalpic contribution due to the solvation energy of the salt ions, and (c) the interfacial width between the two blocks $(N\chi_{\text{eff}})^{-0.5}$. Here, $r$ is the salt loading defined as the ratio of the Li+ and EO monomer concentrations $r=$[Li$^+$]/[EO], and $\chi_{\text{eff}}$ is the effective Flory parameter for the salt-doped block copolymers and is shown to scale as $(\chi_{\text{eff}}-\chi)\propto r$ experimentally [9,37] and theoretically [38]. Mean-field theory suggests that the Born solvation energy, which is proportional to the salt concentration $r$, qualitatively accounts for the experimental data [38]. Accordingly, we set

$$f_1 = r\ln r$$
$$f_2 = ar \qquad (4)$$

according to the Born solvation energy form, and

$$f_3 = (N\chi_{\text{eff}})^{-0.5} \qquad (5)$$

where $\chi_{\text{eff}} = \chi + br$. Both experimental and theoretical studies have suggested that $b$ is on the order of unity; thus, we use $b=1$. The Born solvation energy suggests that $a$ is on the order of 10; thus, we use $a=15$. Here, $a$ varied from $a=1$ to $a=100$, but there was no substantial difference in optimization speed. As in the cases of the polymer solutions and the salt-free diblock copolymer melts, we also examined six different cases for a combination of the $f_i$ (TABLE I), which resulted in significantly poor success rates. Figure 4(a) illustrates our training dataset, consisting of 567 data points constructed from Ref. [9]. The DNN used in Figure 4(b) is tuned with a relative error of 5% and corresponds reasonably well with the experimental data. Although this study examined more than ten other sets of weights with similar accuracy, most of the results compared favorably with the experimental data.

Of particular interest is the predictive power of the DNN when experimental data points are limited. To this end, we consider two limiting cases anticipated from statistical thermodynamics as follows: (1) The phase behavior exhibits DIS in the high-temperature limit, and (2) when the salt concentration is high enough to form an ionic liquid and no additional interaction is considered, the system becomes a dilute block copolymer solution and should exhibit DIS. In the current study, we add these data points to the training datasets as a regularization scheme [39].

We first tune the DNN for only DIS, LAM, and HEX, and not for GYR, with the additional data points from item (1). Figure 4(c) and (d) show our representative results. The DNN predicts GYR around the region at [Li$^+$]/[EO] = 0.05 and T = 130 ºC, as observed in the experiments. Thus, the DNN supports the experimental observation of GYR. This result indicates that even without experimental data for GYR, given information about some other phase behaviors, the DNNs can suggest the location of GYR in the phase diagrams. We suggest that this fact may substantially reduce various experimental costs and may become useful when we explore phases that cannot be easily accessed by experiments and molecular simulations. This study examined more than ten different types of weights with relative errors less than 5%, but all cases predicted GYR near this region. Here, the success rate became 2.03×(%error) – 8.22 and 28.96×(%error -7.5)$^{0.5965}$+7.11 when (%error) ≤ 7.5% and (%error) > 7.5%, respectively. The slope of the success rate was discontinuous at 7.5%. Similarly, using the additional data points from item (2), the DNN predicts that HEX occurs as the salt concentration increases. This result, including the overall trend of the phase boundary of HEX, also corresponds well with the experimental data shown in Figure 4 (a). Here, nineteen out of twenty-three different sets of weights with relative errors less than 5% predicted HEX with reasonable salt concentrations and temperatures. We attribute these reasonable, robust predictions regarding GYR and HEX primarily to the addition of the data points from the limiting cases. For cross validation, we randomly obtained the training data points again, but the overall conclusion remained unchanged. Note that in general, the predicted overall phase diagrams may become patchy and hence qualitatively unacceptable even when training datasets precisely agree with test datasets. This unexpected discrepancy is known in statistics as overfitting, but we have not encountered this problem in this study. Thus, the current DNNs are robust enough to draw the phase diagrams.

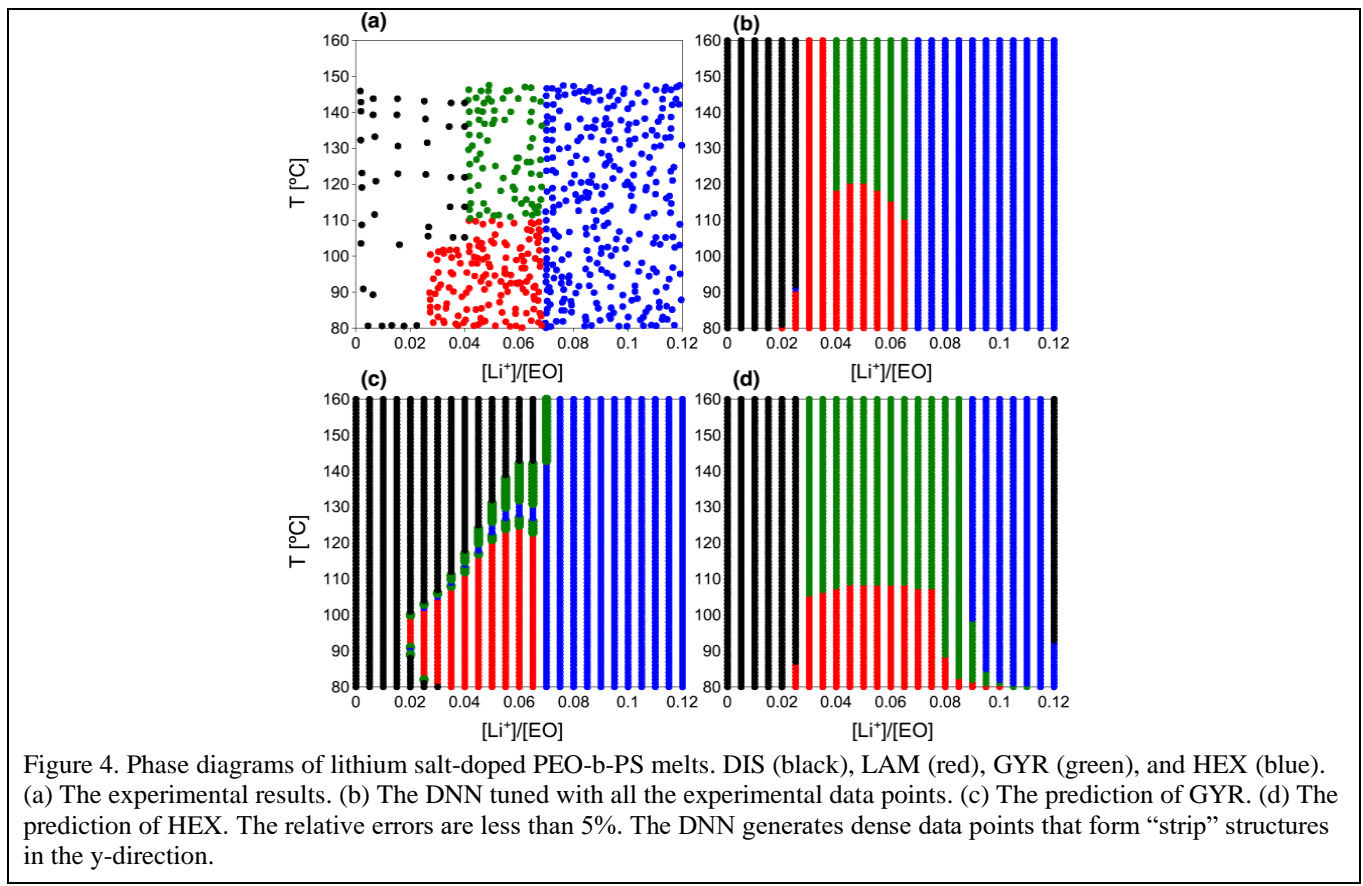

Figure 4. Phase diagrams of lithium salt-doped PEO-b-PS melts. DIS (black), LAM (red), GYR (green), and HEX (blue). (a) The experimental results. (b) The DNN tuned with all the experimental data points. (c) The prediction of GYR. (d) The prediction of HEX. The relative errors are less than 5%. The DNN generates dense data points that form "strip" structures in the y-direction.

## 5. Conclusions

In conclusion, we have developed DNNs that enable drawing the phase diagrams of polymer-containing liquids when the volume fraction (or density) of species, chain length, and the Flory-Huggins parameter $\chi$ (or temperature) are provided. The present study aimed to provide a computationally tractable method that (1) enables studying the phase behaviors that existing molecular theory and simulation cannot easily access and (2) approximates existing (statistical) thermodynamic models while retaining underlying thermodynamic features in targeted systems. A theory-embedded layer that captures thermodynamically important features via mean-field theory and scaling laws was incorporated into a DNN, thus allowing us to describe the phase diagrams of polymer solutions and salt-free and salt-doped diblock copolymer melts. Appropriate construction of this layer significantly speeds up weight optimization and reduces the size of the DNNs, whereas the layer design is not unique. Thus, the design of the theory-embedded layer may become heuristic when applied to other problems. The present study focused on the feedforward network and ReLU for activation functions to consider the most straightforward network systems for conciseness. The effect of the theory-embedded layer on the network performance of other neural network architectures should also be of further interest in future studies. Given that the DNN is tuned, the phase diagrams can be reproduced without implementing advanced theories and molecular simulations. Moreover, our DNN has predictive power and can assist in exploring new phase behaviors in polymer-containing liquid mixtures. Finally, we expect that the current DNNs can be extended to study inverse problems in soft-matter physics [40] that cannot be easily addressed by existing theories and molecular simulations. Specifically, it would be interesting to examine whether the DNNs can predict the chain length and composition of polymers when a particular phase at a specific temperature is given.

## Acknowledgements

The author acknowledges Dr. Jeff Z. Y. Chen at the University of Waterloo and Dr. Mark. J. Stevens at Sandia National Laboratories for providing valuable suggestions and comments on this manuscript. The author is grateful to the High-Performance Computing Shared Facility, Superior, at MTU for their essential support. This work was supported by the Research Excellence Fund of Michigan Technological University.

## References

[1] G. H. Fredrickson, *The equilibrium theory of inhomogeneous polymers* (Oxford University Press, 2006).
[2] J.-P. Hansen and I. R. McDonald, *Theory of simple liquids : with applications of soft matter* (Elsevier / Academic Press, Amsterdam, 2013), Fourth edition. edn.
[3] I. Nakamura and Z.-G. Wang, ACS Macro Lett. **3**, 708 (2014).
[4] I. Nakamura, J. C. Shock, L. Eggart, and T. Gao, Isr J Chem **58**, 1 (2018).
[5] S. Shojaei-Zadeh, J. F. Morris, A. Couzis, and C. Maldarelli, J Colloid Interf Sci **363**, 25 (2011).
[6] T. Kietzke, D. Neher, M. Kumke, O. Ghazy, U. Ziener, and K. Landfester, Small **3**, 1041 (2007).
[7] R. B. Thompson, V. V. Ginzburg, M. W. Matsen, and A. C. Balazs, Science **292**, 2469 (2001).
[8] J. M. Virgili, M. L. Hoarfrost, and R. A. Segalman, Macromolecules **43**, 5417 (2010).
[9] N. S. Wanakule, J. M. Virgili, A. A. Teran, Z.-G. Wang, and N. P. Balsara, Macromolecules **43**, 8282 (2010).
[10] P. M. Simone and T. P. Lodge, ACS Appl. Mater. Interfaces **1**, 2812 (2009).
[11] Y. LeCun, Y. Bengio, and G. Hinton, Nature **521**, 436 (2015).
[12] M. S. Jorgensen, H. L. Mortensen, S. A. Meldgaard, E. L. Kolsbjerg, T. L. Jacobsen, K. H. Sorensen, and B. Hammer, J. Chem. Phys. **151**, 054111 (2019).
[13] S. Ibric, M. Jovanovic, Z. Djuric, J. Parojcic, L. Solomun, and B. Lucic, J Pharm Pharmacol **59**, 745 (2007).
[14] T. Degim, J. Hadgraft, S. Ilbasmis, and Y. Ozkan, J Pharm Sci **92**, 656 (2003).
[15] G. Pilania, C. C. Wang, X. Jiang, S. Rajasekaran, and R. Ramprasad, Sci Rep-UK **3** (2013).
[16] B. S. Rem, N. Käming, M. Tarnowski, L. Asteria, Nick Fläschner, C. Becker, K. Sengstock, and C. Weitenberg, Nat Phys **15**, 917 (2019).
[17] E. P. L. van Nieuwenburg, Y. H. Liu, and S. D. Huber, Nat Phys **13**, 435 (2017).
[18] C. D. Li, D. R. Tan, and F. J. Jiang, Ann Phys-New York **391**, 312 (2018).
[19] Q. S. Wei, R. G. Melko, and J. Z. Y. Chen, Physical Review E **95** (2017).
[20] M. Gao, L. T. Yin, and J. C. Ning, Atmos Environ **184**, 129 (2018).
[21] A. Esteva, B. Kuprel, R. A. Novoa, J. Ko, S. M. Swetter, H. M. Blau, and S. Thrun, Nature **542**, 115 (2017).
[22] T. J. Brinker *et al.*, J Med Internet Res **20** (2018).
[23] S. Dai, B. G. Sumpter, and D. W. Noid, J Phase Equilib **16**, 493 (1995).
[24] C. J. Richardson, A. Mbanefo, R. Aboofazeli, M. J. Lawrence, and D. J. Barlow, J Colloid Interf Sci **187**, 296 (1997).
[25] L. Djekic, S. Ibric, and M. Primorac, International Journal of Pharmaceutics **361**, 41 (2008).
[26] S. Agatonovic-Kustrin and R. G. Alany, Pharmaceut Res **18**, 1049 (2001).
[27] S. Agatonovic-Kustrin, B. D. Glass, M. H. Wisch, and R. G. Alany, Pharmaceut Res **20**, 1760 (2003).
[28] R. G. Alany, S. Agatonovic-Kustrin, T. Rades, and I. G. Tucker, J Pharmaceut Biomed **19**, 443 (1999).
[29] A. Mendyk and R. Jachowicz, Expert Syst Appl **32**, 1124 (2007).
[30] S. Agatonovic-Kustrin, D. W. Morton, and R. Singh, Colloid Surface A **415**, 59 (2012).
[31] J. Bergstra and Y. Bengio, J Mach Learn Res **13**, 281 (2012).
[32] M. Doi, *Introduction to polymer physics* (Clarendon, 1996, Oxford, 1995).
[33] M. D. Rubinstein and R. H. Colby, *Polymer physics* (Oxford University Press, Oxford ; New York, N.Y., 2003).

[34] E. Helfand and Y. Tagami, J. Chem. Phys. **56**, 3592 (1972).
[35] L. Leibler, Macromolecules **13**, 1602 (1980).
[36] I. W. Hamley and V. E. Podneks, Macromolecules **30**, 3701 (1997).
[37] W. S. Young, J. N. L. Albert, A. B. Schantz, and T. H. Epps, Macromolecules **44**, 8116 (2011).
[38] I. Nakamura, N. P. Balsara, and Z.-G. Wang, Phys. Rev. Lett. **107**, 198301 (2011).
[39] J. Schmidhuber, Neural Networks **61**, 85 (2015).
[40] J. F. Li, H. D. Mang, and J. Z. Y. Chen, Phys. Rev. Lett. **123** (2019).